\documentclass[
    prx, 
    reprint,           
    superscriptaddress,
    amssymb,         
    floatfix，
    amsmath,
    aps,
]{revtex4-2}

\usepackage{graphicx}% Include figure files
\usepackage{dcolumn}% Align table columns on decimal point
\usepackage{bm}% bold math
\usepackage{hyperref}
\usepackage{float}
\usepackage{amsmath}  %%% 公式排版必须加载
\usepackage{siunitx} % 物理单位支持 (国际标准格式)
\usepackage{mhchem} % Chemical formula typesetting
\usepackage[utf8]{inputenc}
\usepackage{textgreek}

\usepackage[english,main=english]{babel}
\usepackage{newunicodechar}
\newunicodechar{ }{\,}

\begin{document}

\preprint{APS/123-QED}

\title{Probing Anharmonic Lattice Dynamics and Thermal Transport in Layered Perovskite $\rm LiYTiO_4$ Anode}% Force line breaks with \\

\author{Lin Zhang}
\affiliation{College of Physics, Chongqing University, Chongqing, China}
\affiliation{Center of Quantum Materials and Devices, Chongqing University, Chongqing, China}

\author{Wen Liu}
\affiliation{College of Physics, Chongqing University, Chongqing, China}
\affiliation{Center of Quantum Materials and Devices, Chongqing University, Chongqing, China}

\author{Mingquan He}
\email{mingquan.he@cqu.edu.cn}
\affiliation{College of Physics, Chongqing University, Chongqing, China}
\affiliation{Center of Quantum Materials and Devices, Chongqing University, Chongqing, China}

\author{Jun Huang}%
\email{huangjun2003@126.com}
\affiliation{School of Chemistry and Chemical Engineering, Frontiers Science Center for Transformative Molecules, Shanghai Jiao Tong University, Shanghai 200240, China}

\author{Xiaolong Yang}%
\email{yangxl@cqu.edu.cn}
\affiliation{College of Physics, Chongqing University, Chongqing, China}
\affiliation{Center of Quantum Materials and Devices, Chongqing University, Chongqing, China}

\date{\today}% It is always \today, today,
%\affiliation{%


%\affiliation{
%\affiliation{
%\affiliation{%
%             %  but any date may be explicitly specified

\begin{abstract}
Layered perovskite lithium yttrium titanate ($\rm LiYTiO_4$) has recently emerged as a promising low‑potential, ultrahigh‑rate intercalation‑type anode material for lithium‑ion batteries; however, its lattice dynamics and thermal transport properties remain poorly understood, limiting a complete evaluation of its practical potential. Here, we combine experimental measurements with theoretical modeling to systematically investigate the anharmonic lattice dynamics and heat transport in $\rm LiYTiO_4$. We employ a neural evolution potential (NEP)-based framework that integrates the temperature-dependent effective potential method with the Wigner thermal transport (WTT) formalism, explicitly including both diagonal and off-diagonal terms of the heat-flux operator. Zero‑temperature phonon calculations reveal dynamical instabilities associated with $\rm TiO_6$ octahedral rotation, which are stabilized at finite temperatures through anharmonic renormalization. Using the WTT approach with contributions from phonon propagation and coherence contributions, we predict a room-temperature lattice thermal conductivity ($\kappa_{\rm L}$) of 3.8 $\rm Wm^{-1}K^{-1}$ averaged over all crystal orientations, in close agreement with the measured value of 3.2 ± 0.08 $\rm Wm^{-1}K^{-1}$ for polycrystalline samples. To further examine the possible influence of ionic motion on high‑temperature thermal transport, we compute $\kappa_{\rm L}$ using a Green–Kubo equilibrium molecular dynamics approach based on the same NEP, which yields consistent results with both experiment and WTT predictions, confirming the negligible role of Li‑ion mobility in heat conduction. Our study not only identifies the ultralow thermal conductivity of $\rm LiYTiO_4$ as a key limitation for its practical application but also establishes a reliable computational framework for studying thermal properties in battery materials.

%\begin{description}
%\end{description}
\end{abstract}

\maketitle

%\tableofcontents

\section{\label{sec:level1}introduction}

%In conventional batteries with liquid electrolytes, heat generation during charging and discharging is inevitable. This heat primarily originates from Joule heating induced by internal resistance, entropic contributions from electrochemical reactions, and polarization losses \cite{maher_understanding_2024, noelle_internal_2018}. If not efficiently dissipated, the accumulated heat can result in uneven temperature distribution, accelerated material degradation, and reduced cycle life. In extreme scenarios, it may even trigger thermal runaway, posing serious safety risks \cite{PRXEnergy.1.031002, ZHANG2022121652, offer_cool_2020}. In this context, establishing a comprehensive understanding of thermal transport properties is crucial for the practical application of battery materials\cite{10.1063/5.0013716}. 

Lithium-ion batteries (LIBs) stand as the cornerstone of modern energy storage technology, enabling widespread applications including portable electronic devices and electric vehicles \cite{Zhang2019, Liu2018, Schmuch2018, Li2020}. However, the ongoing pursuit of higher energy density, power density, and fast-charging capabilities has led to significant Joule heating during high-rate cycling \cite{offer_cool_2020, ZHANG2022121652}, posing a fundamental challenge to both the performance and safety of batteries. This internal heat generates from multiple sources: internal resistance, entropic contributions of electrochemical reactions, and polarization losses \cite{maher_understanding_2024, noelle_internal_2018}. If not efficiently dissipated, accumulated heat can give rise to uneven temperature distribution, accelerated material degradation, shortened cycle life, and in severe cases, thermal runaway \cite{PRXEnergy.1.031002, ZHANG2022121652, offer_cool_2020}. Hence, effective thermal management is essential to ensure the overall performance, safety, and operational reliability of battery devices. 

The efficacy of battery thermal management is fundamentally dictated by the intrinsic thermal transport performance of its constituent components, particularly the active electrode materials \cite{PRXEnergy.1.031002}. The low lattice thermal conductivity ($\kappa_L$) of electrode materials cause localized overheating, which not only accelerates the performance degradation but also raises serious safety risks \cite{https://doi.org/10.1002/adfm.202214501}. Nevertheless, current research on electrode materials primarily concentrates on optimizing electrochemical metrics such as specific capacity, rate performance and cycle stability, while systematic investigation into their fundamental thermophysical properties, especially thermal conductivity, remains largely insufficient. This knowledge gap substantially hinders accurate prediction of the practical application potential and life span of electrode materials under realistic operating conditions. Recently, sub-micrometer lithium yttrium titanate ($\rm LiYTiO_4$) with a layered perovskite structure has emerged as a promising low-potential and ultrahigh-rate intercalation-type anode for high-performance LIBs \cite{https://doi.org/10.1002/aenm.202200922}. Although its electrochemical performance has been well established, the underlying lattice dynamics and thermal transport behavior remain elusive, which severely limits a complete assessment of its practical viability and application prospects. 

Electrode materials such as $\rm LiYTiO_4$ are typically multi-elements compounds with complex crystal structures \cite{unnamed, PRXEnergy.4.013012}. Their unit cells contain a large number of atoms interconnected through hierarchical bonding, which gives rise to a dense phonon band structure and strong anharmonicity \cite{doi:10.1021/acs.jpclett.2c03061, Niedziela2019, 10.1021_acs.jpclett.2c00904}. These features make the theoretical prediction of thermal transport in these materials rather challenging. The conventional Boltzmann transport equation (BTE), based on the assumption of well-separated phonon branches, captures only particle-like phonon propagation but neglects the wave-like tunneling between phonon eigenstates \cite{xia_particlelike_2020, PhysRevLett.62.645}. As a result, it often fails to fully explain the thermal conductivity behavior in such systems \cite{PhysRevB.105.224303, https://doi.org/10.1002/smll.202506386}. To overcome this shortcoming, many efforts have been made to establish unified theories and numerical techniques beyond the conventional BTE scheme. Notable progress in thermal transport theory has now enabled explicit consideration of the wave-particle duality of phonons \cite{doi:10.1126/science.aar8072, 10.1038_s41567-019-0520-x, 10.1038_s41467-019-11572-4, 10.1038_s41467-020-16371-w}. Among these advances, Simoncelli et {\em al.} developed a unified theory by reformulating the BTE within the Wigner formalism \cite{simoncelli_unified_2019, Simoncelli2021WignerFO}, using a phonon velocity operator whose diagonal and off-diagonal elements respectively describe propagative and diffusive contributions to heat transport. This approach has been successfully applied to evaluate the $\kappa$ of numerous strongly anharmonic and structurally complex crystals \cite{xia_microscopic_2020,  10.1038_s41524-024-01210-z, 10.1093/nsr/nwae216, doi:10.1021/acs.nanolett.3c02957}, thereby providing a solid theoretical foundation for studying thermal transport in battery materials. Nevertheless, it should be noted that the Wigner formalism, being rooted in lattice dynamics, does not account for the effects of potential Li-ion migration, as observed in superionic systems such as $\rm Ag_9GaSe_6$ \cite{LIN2017816}, $\rm Ag_8SnSe_6$ \cite{10.1038_s41563-023-01560-x}, and $\rm CsCu_2I_3$ \cite{https://doi.org/10.1002/adma.202513381}, which may exert a non-negligible influence on $\kappa$. In contrast, molecular dynamics (MD) simulations inherently capture such ion migration along with wave-like tunneling and higher-order anharmonicity \cite{10.1103/PhysRevB.105.115202, 10.1038_s41524-022-00776-w}, offering a powerful alternative for investigating thermal transport in LIB materials.

In this work, we present a systematic study of the anharmonic lattice dynamics and thermal transport behavior in $\rm LiYTiO_4$, combining experimental measurements with theoretical modeling. We employ a machine-learning potential-based framework that incorporates the temperature-dependent effective potential (TDEP) method together with the Wigner thermal transport (WTT) formalism, accounting for both three-phonon (3ph) and four-phonon (4ph) scatterings within the diagonal and non-diagonal terms of the heat flux operator. Phonon calculations at 0 K reveal dynamical instabilities associated with opposite rotations of the adjacent $\rm TiO_6$ octahedra. Through anharmonic renormalization, these strongly unstable optical modes are stabilized at finite temperatures. We then compute the lattice thermal conductivity using the WTT approach, incorporating both particle-like and wave-like coherence contributions. The predicted room-temperature (RT) average $\kappa_L$ is $3.8\ \rm Wm^{-1}K^{-1}$, in good agreement with the experimental value of $3.2 \pm 0.08\ \rm Wm^{-1}K^{-1}$. To assess the influence of ion hopping on high-temperature thermal transport, we also evaluate the $\kappa_L$ using the neural evolution potential (NEP)-based Green–Kubo equilibrium molecular dynamics (GK-EMD) method. The results closely match both the experimental data and the WTT predictions, indicating a negligible role of Li-ion mobility in heat conduction. Our findings not only clarify the limited role of ionic diffusion in thermal transport but also establish a reliable computational framework for modeling thermal transport properties in battery materials.

\begin{figure}[t]
    \centering
    \includegraphics[width=1.0\linewidth]{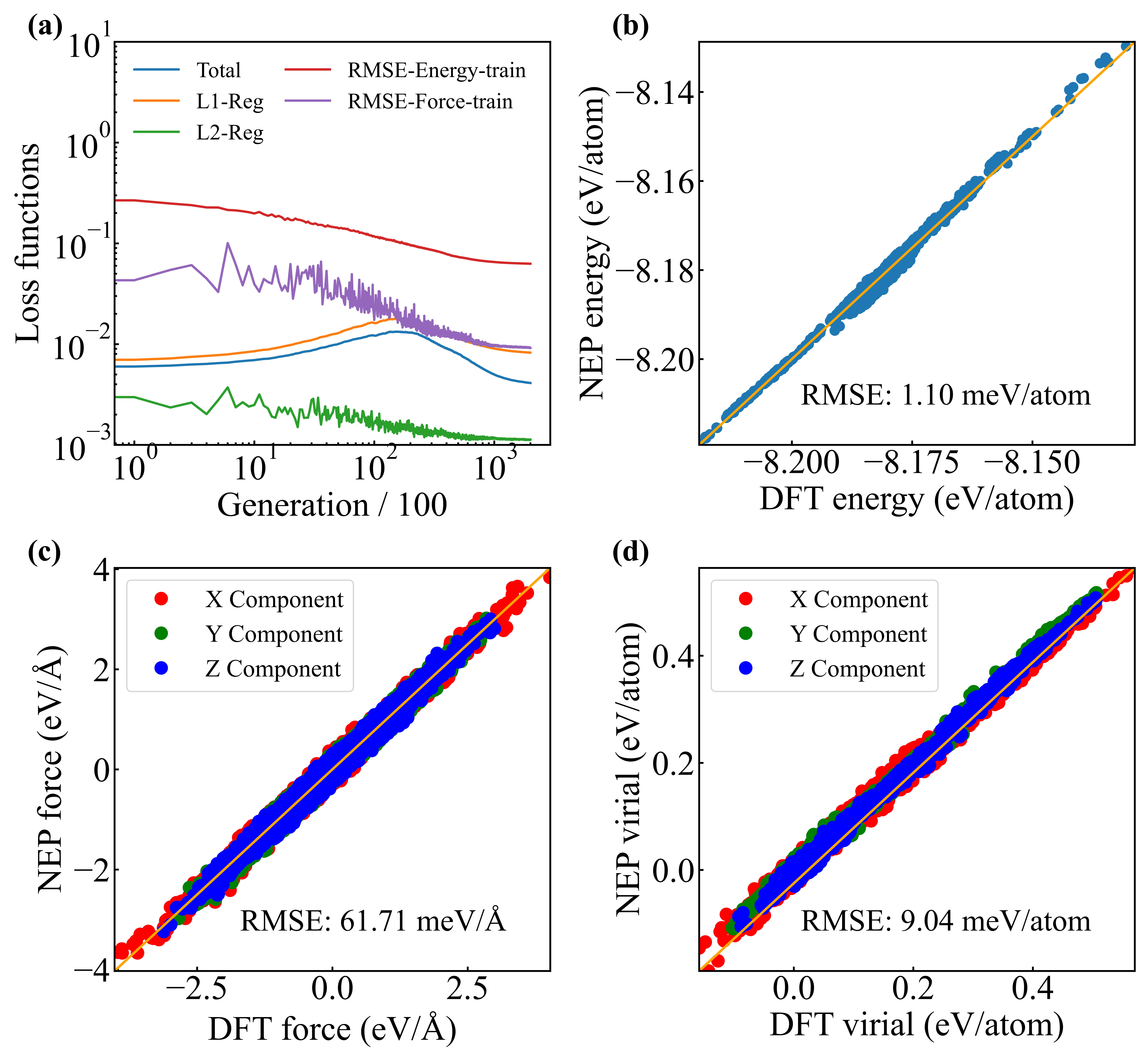}  
    \caption{Training results showing the convergence of energy, forces, stresses, and error functions. The high consistency of the diagonal correlation plot indicates the strong predictive capability of the NEP.}
    \label{fig:training_results}
\end{figure}

\section{METHODOLOGY}

\subsection{Neuroevolution machine learning potential}
To navigate the trade-off between computational accuracy and cost, we employed a neuroevolution machine learning potential (NEP) to describe interatomic interactions in $\rm LiYTiO_4$. The training and testing database was constructed using first-principles molecular dynamics simulations based on density functional theory (DFT). Structural optimizations and ab initio molecular dynamics (AIMD) simulations were carried out using the VASP \cite{PhysRevB.54.11169}, employing the projector-augmented wave (PAW) method \cite{PhysRevB.50.17953}. The exchange–correlation functional was treated within the generalized gradient approximation (GGA) using the Perdew–Burke–Ernzerhof (PBE) formulation \cite{PhysRevLett.77.3865}. For structural optimization of the unit cell, a $2 \times 2 \times 2$ Monkhorst–Pack $\bf k$-point mesh \cite{PhysRevB.13.5188} was used for Brillouin zone integration, and a plane-wave cutoff energy of 550 eV was applied. The convergence criteria for the electronic self-consistent iteration and ionic relaxation were set to $10^{-8}$ eV for energy and $0.01$ eV/Å for forces, respectively. The optimized lattice parameters obtained from DFT are $a = 11.27$ Å, $b = 5.46$ Å, and $c = 5.35$ Å, which agree well with the experimental values of $a = 11.0126$ Å, $b = 5.3880$ Å, and $c = 5.3870$ Å \cite{https://doi.org/10.1002/aenm.202200922}. AIMD simulations were carried out on a $1 \times 2 \times 2$ supercell (112 atoms) in the canonical (NVT) ensemble to sample energies, forces, and stresses across atomic configurations. Temperatures ranging from 100 to 800 K were sampled with a time step of 1 fs, using a Nosé-Hoover thermostat \cite{A-Unified-Formulation-of-the-Constant-Temperature-Molecular-Dynamics-Method} for temperature control. To enhance structural diversity, additional configurations were collected at 300 K under the NPT ensemble. Furthermore, to improve the transferability of the NEP, we applied $\pm 2\%$ strains to the original configurations, generating four deformed structures. For each deformed structure, atomic trajectories were generated using the same AIMD settings. All AIMD simulations employed a $\Gamma$-centered $2 \times 2 \times 2$ $\bf k$-point grid. 

To reduce frame correlation and account for the data efficiency of the NEP model, one structure was extracted every 10 steps from the AIMD trajectories, yielding an initial set of 574 structures for the first training round. The training set was subsequently expanded through an active learning strategy: from the remaining data, the MaxVol algorithm \cite{podryabinkin_active_2017} was used to select 500 configurations that were structurally most distinct from the initial training set. The NEP obtained from the first training round was used to evaluate the energies and forces of these configurations, which were then incorporated into the training set. This resulted in a final training set of 1074 structures for the second round of training. During training, the radial and angular descriptor cutoffs were set to 8.0 $\text{\AA}$ and 4.0 $\text{\AA}$, respectively, and the model quality was evaluated using the root-mean-square error (RMSE). To validate the accuracy of the trained NEP, we compared the phonon spectra derived from the NEP with those computed directly from DFT. The harmonic interatomic force constants (IFCs) at the ground state (0 K) were calculated using the finite displacement method with a $2 \times 2 \times 2$ supercell, as implemented in the ALAMODE package \cite{tadano_anharmonic_2014}.

\subsection{Thermal conductivity based on Wigner formalism}
To accurately predict the lattice thermal conductivity of  $\rm LiYTiO_4$, we employ the linearized Wigner thermal transport (WTT) equation that explicitly includes the contributions from particle-like phonon propagation ($\kappa_{\mathrm{L}}^{\rm P}$) and wave-like tunneling of phonons ($\kappa_{\mathrm{L}}^{C}$). Under the single-mode relaxation time approximation (SMRTA), the equation is formulated as \cite{xia_microscopic_2020, simoncelli_unified_2019, feng_quantum_2016}

\begin{multline}
\kappa_{\mathrm{L}}^{\mathrm{P/C}} = \frac{\hbar^{2}}{k_{\mathrm{B}}T^{2}VN} 
\sum_{\mathbf{q}}\sum_{j,j^{\prime}} \frac{\Omega_{\mathbf{q}j}+\Omega_{\mathbf{q}j^{\prime}}}{2}
\, u_{\mathbf{q}jj^{\prime}} \otimes u_{\mathbf{q}j^{\prime}j} \\
\times \frac{\Omega_{\mathbf{q}j} n_{\mathbf{q}j} (n_{\mathbf{q}j}+1) + \Omega_{\mathbf{q}j^{\prime}} n_{\mathbf{q}j^{\prime}} (n_{\mathbf{q}j^{\prime}}+1)}
{4(\Omega_{\mathbf{q}j}-\Omega_{\mathbf{q}j^{\prime}})^{2} + (\Gamma_{\mathbf{q}j}+\Gamma_{\mathbf{q}j^{\prime}})^{2}} 
(\Gamma_{\mathbf{q}j}+\Gamma_{\mathbf{q}j^{\prime}}),
\end{multline}

where $j$ and $\bf q$ denote the mode index and wave vector of a phonon mode, respectively. $k_{\mathrm{B}}$ is the Boltzmann constant, $T$ the temperature, $V$ the primitive-cell volume, $N$ the total number of $\bf q$-points sampled in the first Brillouin zone, and $n_{\mathbf{q}j^{\prime}}$ the phonon occupation following the Bose–Einstein distribution. $\Omega_{\mathbf{q}j}$ represents the anharmonically renormalized phonon frequency, and $u_{\mathbf{q}j^{\prime}j}$ is the generalized group velocity that involves both intra-branch and inter-branch components \cite{allen_thermal_1993}. In this expression, the diagonal terms ($j=j'$) of the heat flux operator correspond to the populations contribution ($\kappa_{\mathrm{L}}^{\mathrm{P}}$), whereas the off-diagonal terms ($j \ne j'$) of the heat flux operator yield the coherences' contribution ($\kappa_{\mathrm{L}}^C$). The total lattice thermal conductivity is given by the sum of these two components: $\kappa_{\mathrm{L}} = \kappa_{\mathrm{L}}^{\mathrm{P}} + \kappa_{\mathrm{L}}^{\mathrm{C}}$. Here, the $\Gamma_{\mathbf{q}j}$ stands for the total scattering rate, accounting for scattering from three-phonon (3ph) and four-phonon (4ph) interactions, as well as isotope scattering. These rates were computed based on Fermi’s golden rule \cite{feng_quantum_2016} using the temperature-dependent effective potential (TDEP) approach with renormalized interatomic force constants (IFCs) up to fourth order (harmonic, third- and fourth-order anharmonic terms). The individual linewidths due to 3ph and 4ph scattering processes are given by~\cite{feng_quantum_2016, xia_particlelike_2020, xia_microscopic_2020}

\begin{flalign}
\Gamma_{q}^{\mathrm{3ph}} = \sum_{q^{\prime}q^{\prime\prime}} \Bigg\{ 
& \frac{1}{2}\left(1+n_{q^{\prime}}^{0}+n_{q^{\prime\prime}}^{0}\right)\Lambda_{-} \notag \\
& + \left(n_{q^{\prime}}^{0}-n_{q^{\prime\prime}}^{0}\right)\Lambda_{+}
\Bigg\}, &
\end{flalign}

\begin{flalign}
\Gamma_{q}^{\mathrm{4ph}}
= \sum_{q^{\prime}q^{\prime\prime}q^{\prime\prime\prime}} \Bigg\{ 
& \frac{1}{6}\frac{n_{q^{\prime}}^{0}n_{q^{\prime\prime}}^{0}n_{q^{\prime\prime\prime}}^{0}}{n_{q}^{0}}\Lambda_{--} \notag \\
& + \frac{1}{2}\frac{\left(1+n_{q^{\prime}}^{0}\right)n_{q^{\prime\prime}}^{0}n_{q^{\prime\prime\prime}}^{0}}{n_{q}^{0}}\Lambda_{+-} \notag \\
& + \frac{1}{2}\frac{\left(1+n_{q^{\prime}}^{0}\right)\left(1+n_{q^{\prime\prime}}^{0}\right)n_{q^{\prime\prime\prime}}^{0}}{n_{q}^{0}}\Lambda_{++}
\Bigg\}, &
\end{flalign}

\begin{flushleft}
where $\Lambda_{\pm}$ and $\Lambda_{\pm\pm}$ are defined as
\end{flushleft}

\begin{equation}
\Lambda_{\pm} = \frac{\pi}{4N} 
\left| V^{(3)} \left( q, \pm q', -q'' \right) \right|^{2} 
\Delta_{\pm} 
\frac{\delta \!\left( \Omega_{q} \pm \Omega_{q'} - \Omega_{q''} \right)}
{\Omega_{q}\Omega_{q'}\Omega_{q''}} ,
\end{equation}

\begin{flalign}
&\Lambda_{\pm\pm} = 
\frac{\pi^{2}}{8N^{2}}
\left| V^{(4)} \left( q, \pm q', \pm q'', -q''' \right) \right|^{2}
\Delta_{\pm\pm} && \nonumber\\
&\hspace{5em} \times
\frac{\delta \!\left( \Omega_{q} \pm \Omega_{q'} \pm \Omega_{q''} - \Omega_{q'''} \right)}
{\Omega_{q}\Omega_{q'}\Omega_{q''}\Omega_{q'''}}. &&
\end{flalign}

Here the quantities $V^{(3)}(q, \pm q', -q'')$ and $V^{(4)}(q, \pm q', \pm q'', -q''')$ represent the reciprocal-space forms of the third- and fourth-order interatomic force constants (IFCs). The Dirac delta function $\delta(\Omega)$ enforces energy conservation in 3ph and 4ph scattering processes, while the Kronecker symbols $\Delta_\pm$ and $\Delta_{\pm\pm}$, shorthand for $\Delta_{q q' - q''}$ and $\Delta_{q q' q'' - q''', \mathbf{Q}}$, respectively, impose the conservation of crystal momentum.

Solving the WTT equation numerically requires second-, third-, and fourth-order IFCs as inputs, which were obtained using the neural network interatomic potential (NEP) as implemented in the LAMMPS package \cite{PLIMPTON19951}. To incorporate phonon renormalization effects, finite-temperature IFCs were calculated via the TDEP method \cite{Temperature-dependent-effective-potential-method-for-accurate-free-energy-calculations-of-solids}. To reduce the computational cost, the atomic trajectories for TDEP were generated from classical molecular dynamic (MD) simulations using the trained NEP, rather than directly from DFT. At different temperatures, equilibrium MD simulations were performed in the NVT ensemble on a $2 \times 3 \times 3$ supercell containing 504 atoms, running for 100,000 steps with a time step of 1 fs. From these simulations, the renormalized second-, third-, and fourth-order IFCs were extracted using cutoff radii of 4.2, 4.2, and 3 {\AA}, respectively. Thermal conductivity calculations were then carried out using the modified ShengBTE code \cite{ShengBTE_2014, HAN2022108179} with a $15 \times 15 \times 15$ $\bf q$-mesh and a $2 \times 3 \times 3$ supercell, which ensures the convergence of $\kappa_L$, as verified in the Supplementary Material. Given the high computational expense associated with 4ph scattering calculations, we adopted a sampling-estimation approach \cite{10.1038_s41524-023-01020-9}, whereby 4ph scattering rates are estimated from a subset of all possible phonon scattering events. A sample size of $\rm 10^6$ was used for this estimation. 

\subsection{Green-Kubo molecular dynamics}
Once the NEP with a small error is constructed, the lattice thermal conductivity of $\rm LiYTiO_4$ is evaluated using the Green–Kubo equilibrium molecular dynamics (GK-EMD) method as implemented in the open-source GPUMD package \cite{Accelerated-molecular-dynamics-force-evaluation-on-graphics-processing-units-for-thermal-conductivity-calculations, FAN201710}. Based on the Green–Kubo relation \cite{Kubo1985}, the $\kappa_{\rm L}$ is obtained as  
\begin{equation}
\kappa = \frac{1}{3Vk_{\mathrm{B}}T^{2}} \int_{0}^{\infty} \langle \mathbf{J}(t)\cdot\mathbf{J}(0) \rangle  dt,
\end{equation}
where $\bf{J}(t)$ denotes the heat current and the angular bracket represents the autocorrelation function. To mitigate finite-size effects, which in the EMD method are mainly governed by the phonon wavelength, the simulation cell was constructed with a length of approximately 4 nm along the heat-flow direction, corresponding to a $4 \times 8 \times 8$ supercell containing 7168 atoms. The GK-EMD simulations use a time step of 1 fs and periodic boundary conditions in all three directions. A time step of 1 fs was used in the GK-EMD simulations, with periodic boundary conditions applied in all three directions. The system was first equilibrated with a 0.1 ns NVT run, followed by a 1 ns NVE run for stabilization, and then a production run of 1 ns in the NVE ensemble was performed to compute the heat-current autocorrelation function. The maximum correlation time was set to 100 ps, long enough to capture the relevant fluctuation-dissipation behavior. To reduce statistical uncertainties inherent in the GK-EMD approach, 10 independent simulations with different initial velocities were carried out and their results averaged. 
Besides, ionic diffusion was characterized using mean squared displacement (MSD) and radial distribution function (RDF) analyses. The MSD was calculated following the established methodology in Ref.~\cite{Rapaport_2004}. For these analyses, a supercell of 896 atoms was initially equilibrated for 100,000 steps in the NVT ensemble. Subsequently, MSD data were collected over a production run of 1,000,000 steps in the NVE ensemble, while RDF configurations were sampled over 100,000 steps in the NVT ensemble.

%During the simulations, the total simulation time is set to 1000 ps, with the maximum correlation time chosen as 100 ps. The production time is 10 times the maximum correlation time, ensuring a reasonable value. The time step is 1 fs.
%\begin{equation}
%\mathbf{J}(t) = \sum_{i=1}^N \frac{d}{dt}\big(\mathbf{r}_i E_i\big),
%\end{equation}
%where $N$ represents the total number of atoms, $\mathbf{r}_i$ represents the position vector %of atom $i$ and $E_i$ is total energy. Then
%\begin{equation}\mathbf{J(t)}=\sum_{i=1}^N\boldsymbol{v}_iE_i+\sum_{i=1}^N\boldsymbol{r}_i\frac{d}{dt}E_i
%\end{equation}
%In the above equation, the first term represents the kinetic contribution, which plays an important role in fluids, while the second term corresponds to the potential contribution, according to the expression of atomic energy given by

%\begin{equation}
%E_i = \frac{m_i|\mathbf{v}_i|^2}{2} + U_i,
%\end{equation}
%with $m_i$ and $\mathbf{v}_i$ denoting the mass and velocity of atom $i$, and $U_i$ is potential energy. which can be further expressed as
%\begin{equation}\mathbf{J(t)}_{\mathrm{pot}}=\sum_{i=1}^Nr_i(F_i\cdot v_i)+\sum_{i=1}^Nr_i\frac{dU_i}{dt}\end{equation}

%\noindent and the total heat current can thus be written as
%\begin{equation}\mathbf{J(t)}=\sum_{i=1}^N\boldsymbol{v}_iE_i+\sum_{i=1}^Nr_i(F_i\cdot v_i)+\sum_{i=1}^Nr_i\frac{dU_i}{dt}
%\end{equation}

\subsection{Experimental method}
\subsubsection{Synthesis and characterization of $\rm LiYTiO_4$ samples}
This study utilized a sol-gel method coupled with an ion exchange process to synthesize $\rm LiYTiO_4$. The synthesis began with the preparation of a sodium-based precursor, $\rm NaYTiO_4$, via the sol-gel route, followed by a lithium‑sodium ion exchange reaction. Compared to conventional high‑temperature solid‑state calcination, which typically requires prolonged thermal treatment, this approach substantially reduced the calcination time and effectively suppressed high‑temperature‑induced phase separation and abnormal grain growth. Further synthesis details can be found in our earlier publication \cite{https://doi.org/10.1002/aenm.202200922}.

\begin{figure}[htbp]
\centering
\includegraphics[width=1.02\linewidth]{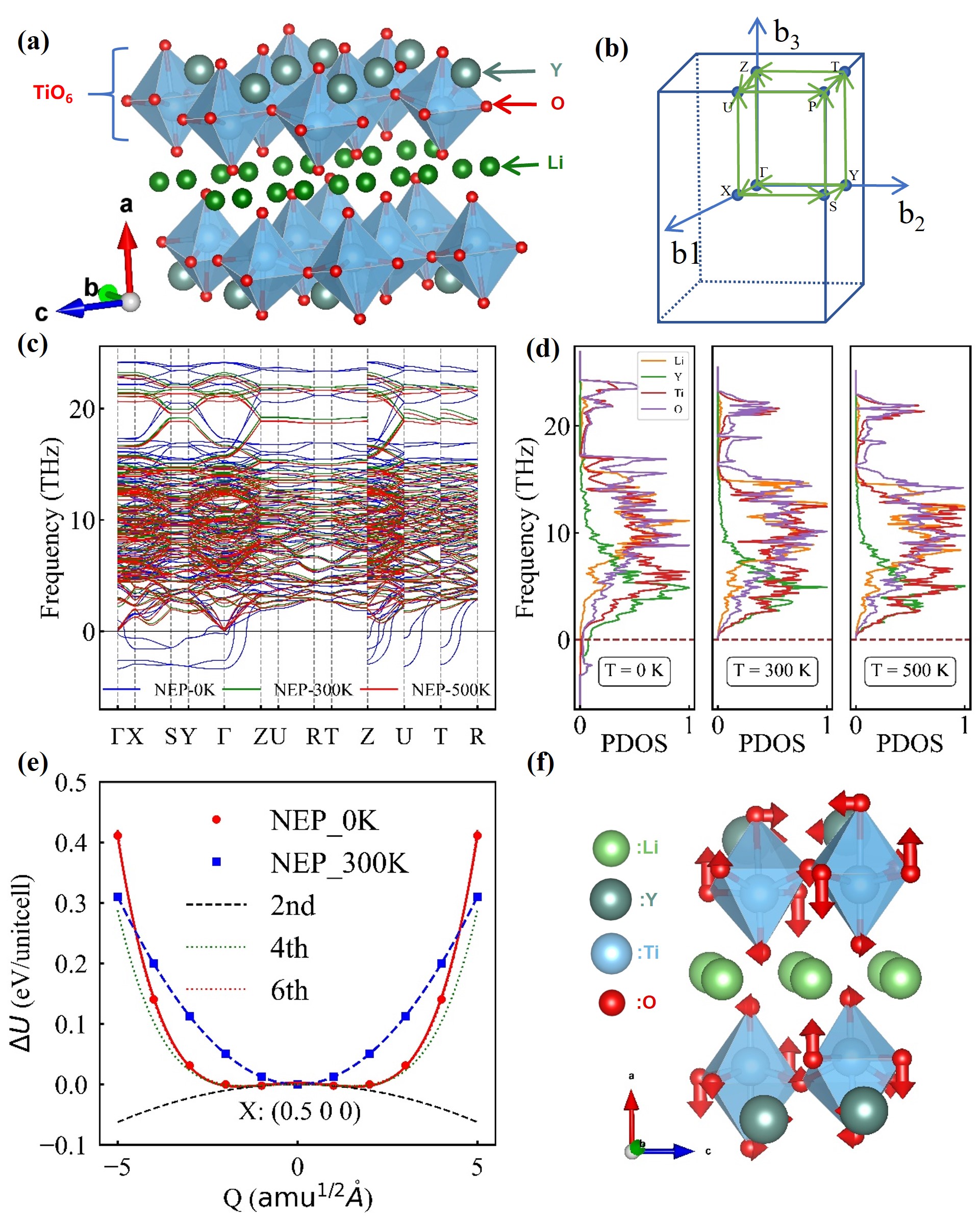} % 适当调整宽度
\caption{
(a) The crystal structure of layered perovskite $\rm LiYTiO_4$ (orthorhombic system, space group Pbcm) features each titanium (Ti) atom is surrounded by six oxygen (O) atoms, forming corner-sharing [$TiO_6$] octahedral units. 
(b) Brillouin zone of $\mathrm{LiYTiO_4}$. 
(c) Comparison of phonon spectra obtained from NEP at 0\,K, 300\,K, and 500\,K. 
(d) Comparison of phonon density of states (PDOS) at 0\,K, 300\,K, and 500\,K. 
(e) Comparison of the potential energy surfaces at (0.5, 0, 0) between the ground state and the TDEP-renormalized at 300 K, dash lines show the potential energy surface decomposed to the second, fourth and sixth orders, respectively. 
(f) The crystal structure of $\mathrm{LiYTiO_4}$ together with the displacement pattern of the soft phonon mode at the X point.
}
\label{fig2}
\end{figure}

$\rm LiYTiO_4$ crystallizes in an orthorhombic layered perovskite structure with space group $Pbcm$ \cite{https://doi.org/10.1002/aenm.202200922}. Refined lattice parameters are $a = 11.0196(5)$ Å, $b = 5.3994(4)$ Å, and $c = 5.3929(5)$ Å, showing in good agreement with our calculations. As shown in Fig.~\ref{fig2}(a), the crystal exhibits a layered arrangement along the $a$‑axis, consisting of alternating layers of $\rm LiO_4$ tetrahedra, $\rm YO_8$ dodecahedra, and $\rm TiO_6$ octahedra. Due to the difficulty in obtaining single crystals, polycrystalline $\rm LiYTiO_4$ samples with grain sizes of approximately 1–2 $\mu$m were synthesized. High‑magnification SEM and TEM images (see supplementary material) show smooth, pore‑ and defect‑free surfaces. High‑resolution TEM reveals clear lattice fringes, and fast Fourier transform (FFT) analysis confirms an interplanar spacing of 0.3663 nm, corresponding to the (300) plane.

\subsubsection{Thermal transport measurements}
\textbf{Compacting powders.} The \ce{LiYTiO4} powders were loaded into a graphite die. The powder samples in the die were densified at \qty{850}{\degreeCelsius} for \qty{12}{\minute} under an axial pressure of \qty{50}{\mega\pascal} in a high vacuum using an SPS system (SPS2000).

\textbf{Thermal conductivity measurements.} A laser flash apparatus (LFA467 HyperFlash) was used to record the thermal transport properties of the samples. The specimens were disc-shaped with a diameter of \qty{9.5}{\milli\meter} and a thickness of \qty{2.956}{\milli\meter}. Both circular faces of the samples were coated with graphite prior to measurement, and the analysis area was \qty{2.8}{\milli\meter} in diameter. The thermal conductivity was calculated from the formula $\kappa_{tot}=DC_{p}\rho$, where $D$ is the thermal diffusivity, $C_{p}$ is the specific heat capacity, and $\rho$ is the mass density of the specimens (\qty{3.62}{\gram\per\centi\meter\cubed}). The $\rho$ value used was obtained by their geometrical dimensions and masses.

% 跨双栏图形

\begin{figure*}[htbp]
\centering
\includegraphics[width=0.7\linewidth]{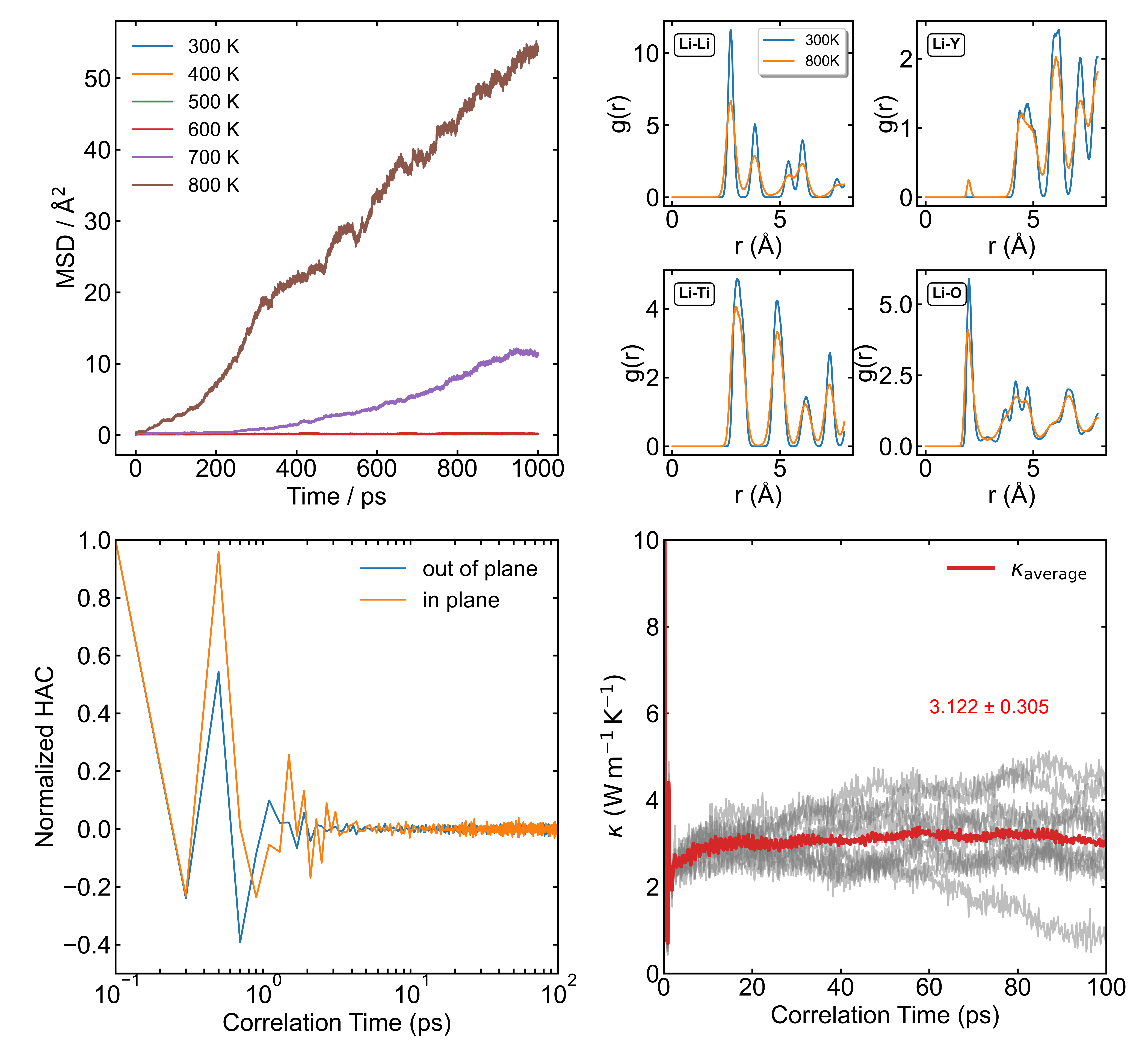} % 适当调整宽度
\caption{
(a)Temperature-dependent mean square displacement (MSD) of Li atoms from 300 to 800\,K. (b) Comparison of RDF at 300 and 800 K. (c)The heat flux autocorrelation function at 300 K. (d)The running thermal conductivity averaged over three crystallographic directions, computed using EMD method at 300 K.
}
\label{fig3}
\end{figure*}

\section{RESULTS AND DISCUSSION}
\subsection{Machine-learning interatomic potentials}
The NEP is trained and tested, and its reliability is summarized in Fig. \ref{fig:training_results}. It is seen that the energy, forces, and stresses, along with the corresponding error metrics, eventually converge to a stable state, indicating excellent generalization ability. Moreover, the diagonal correlation plot exhibits high consistency between predicted and reference values, further confirming the strong predictive power of the NEP. The root-mean-square errors (RMSEs) for energy, forces, and stresses on the test dataset are 1.10, 61.71, 9.04 meV/atom, respectively, demonstrating the high accuracy of the potential. To further validate this potential, we compare the energy–volume curves obtained from both NEP and DFT under 2\% strain, as provided in the Supplementary Materials. The results show good agreement between the two methods, underscoring the reliability of the NEP in reproducing DFT-level potential energy surface. Additionally, the phonon dispersion calculated using the NEP aligns well with the DFT calculations, as shown in the Supplementary Materials. These results collectively demonstrate that our trained NEP possesses high predictive accuracy and robustness, and can be reliably used in subsequent calculations.

\subsection{Anharmonic lattice dynamics}
%As shown in Fig.~\ref{fig2}(a), $\rm LiYTiO_4$ crystallizes in an orthorhombic layered perovskite structure with the space group $Pbcm$ \cite{https://doi.org/10.1002/aenm.202200922}. In this structure, each Ti atom is octahedrally coordinated by six O atoms, forming $\rm TiO_6$ units. These octahedra connect at their corners and are rotated and tilted toward the Li-O layer and away from the Y-O layer, a distortion that yields the orthorhombic symmetry. Li ions occupy tetrahedral sites coordinated by oxygen, forming an antifluorite layer configuration. 
We begin by examining the lattice dynamics of $\rm LiYTiO_4$ within the harmonic approximation (HA) using the NEP. The calculated harmonic phonon dispersions at zero temperature, shown in Fig.~\ref{fig2}(c), reveal several soft phonon branches with imaginary frequencies along high-symmetry directions in the Brillouin zone, indicating dynamical instability of the crystal structure at low temperatures. As evidenced by the atom-projected phonon density of states (PDOS) in Fig.~\ref{fig2}(d), these soft modes arise primarily from vibrations of Y and O atoms. When temperature effects are incorporated via the TDEP technique, the anharmonically renormalized phonon spectra become fully stabilized above room temperature. Apart from the hardening of the low-frequency soft modes with increasing temperature, we find that the high-frequency modes ($>$15 THz), primarily contributed by O and Li atoms, show significant softening, underscoring the strong anharmonicity in the system.

To attain deep insight into the lattice instability and anharmonicity in $\rm LiYTiO_4$, we calculate the PES of the lowest-frequency transverse acoustic (TA) mode at the $\bf X$ point, as shown in Fig.~\ref{fig2}(e). At 0 K, the soft mode clearly exhibits a double-well potential, indicating strong lattice anharmonicity. This anharmonic double-well potential can be accurately reproduced by a polynomial fit that includes up to the sixth-order term. When the PES is decomposed into contributions of different orders, we find that the harmonic term of the soft mode has a negative coefficient, which accounts for the imaginary frequency and thus explains the failure of the conventional HA treatment. Although the PES of the soft mode displays anharmonic behavior up to the sixth order, the contributions from these higher-order terms are relatively minor. Therefore, in our anharmonic phonon renormalization calculations, only anharmonic contributions up to the fourth order are included. When finite-temperature effects are taken into account, the PES at 300 K transforms into a single $U$-shaped well with a positive curvature. This evolution indicates that the collective atomic dynamics, which are unstable at 0 K, become dynamically stable at finite temperatures. Vibrational mode analysis further reveals that the double-well $U$-shaped PES is associated with opposite rotations of adjacent $\rm TiO_6$ octahedra along the crystallographic b-axis, as illustrated in Fig.~\ref{fig2}(f).  

\subsection{Diffusion dynamics of Li ions}
We have previously demonstrated using nudged elastic band calculations that Li ions in pristine $\rm LiYTiO_4$ migrate via a two-dimensional pathway within the $b$–$c$ plane \cite{https://doi.org/10.1002/aenm.202200922}. Such Li-ion diffusion may lead to the breakdown of the phonon picture and potentially induce pronounced phonon anharmonicity, which in turn may profoundly affect thermal transport behavior. It is therefore necessary to conduct an in-depth investigation of Li-ion diffusion dynamics at finite temperatures using machine-learning potential-based MD simulations. The Li ions hopping behavior can be characterized using time-dependent mean square displacement (MSD) analysis, as shown in Fig.~\ref{fig3}(a). At 300 K, the MSD of Li ions remains nearly constant over time, suggesting localized oscillatory motion around their equilibrium positions. However, as temperature rises to 700 K, Li ions begin to migrate, and their MSD increases nearly linearly with time, indicating frequent intralayer hopping between adjacent $\rm TiO_6$ octahedral units. Alongside Li ions, O atoms also exhibit significant diffusion at high temperatures, as revealed by the time-dependent MSD presented in the Supplementary Material. To further corroborate the Li-ion hopping behavior, we also compute the radial distribution functions $g(r)$ of relevant atomic pairs, which allow us to monitor the evolution of the local structure surrounding the Li ions. By comparing the cases at 300 K and 800 K in Fig.~\ref{fig3}(c), distinct changes in the peak width and shape are observed in the $g(r)$ for different atomic pairs. In particular, the $g(r)$ peaks for $\rm Li-Li$ and $\rm Li-O$ pairs become notably lower and broader at elevated temperatures. These results confirm that Li-ion diffusion induces local structural relaxation at high temperatures.

This ion diffusion behavior may render the conventional phonon gas model inadequate or inaccurate for describing thermal transport in $\rm LiYTiO_4$. To overcome this limitation, we employ the NEP-based GK-EMD method, which provides a more comprehensive and accurate prediction of $\kappa_{\mathrm{L}}$ by inherently incorporating all-order anharmonic phonon interactions and explicitly accounting for ionic hopping effects. Figure~\ref{fig3}(c) showcases the normalized heat-flux autocorrelation versus correlation time for the in-plane and out-of-plane directions at 300 K. The results indicate that a truncation time of 10 ps is sufficient to achieve well-converged values. At elevated temperatures, where phonon lifetimes are shorter, even faster convergence of the heat-flux autocorrelation is expected. Accordingly, Fig.~\ref{fig3}(d) presents the calculated $\kappa_{\mathrm{L}}$ as a function of the truncation time, confirming that 10 ps adequately ensures convergence of the thermal conductivity. Based on ten independent EMD simulations, we obtain a statistically averaged thermal conductivity of $3.1 \pm 0.3$ Wm$^{-1}$K$^{-1}$ at 300 K, which agrees well with our measured value of $3.2 \pm 0.1$ Wm$^{-1}$K$^{-1}$. This agreement validates the accuracy of our NEP and GK-EMD simulations. Using the same computational setup, we further calculate the thermal conductivity in the temperature range of 300-800 K to enable a systematic comparison with experimental measurements in the subsequent section.

\begin{figure*}[htbp]
\centering
\includegraphics[width=0.7\linewidth]{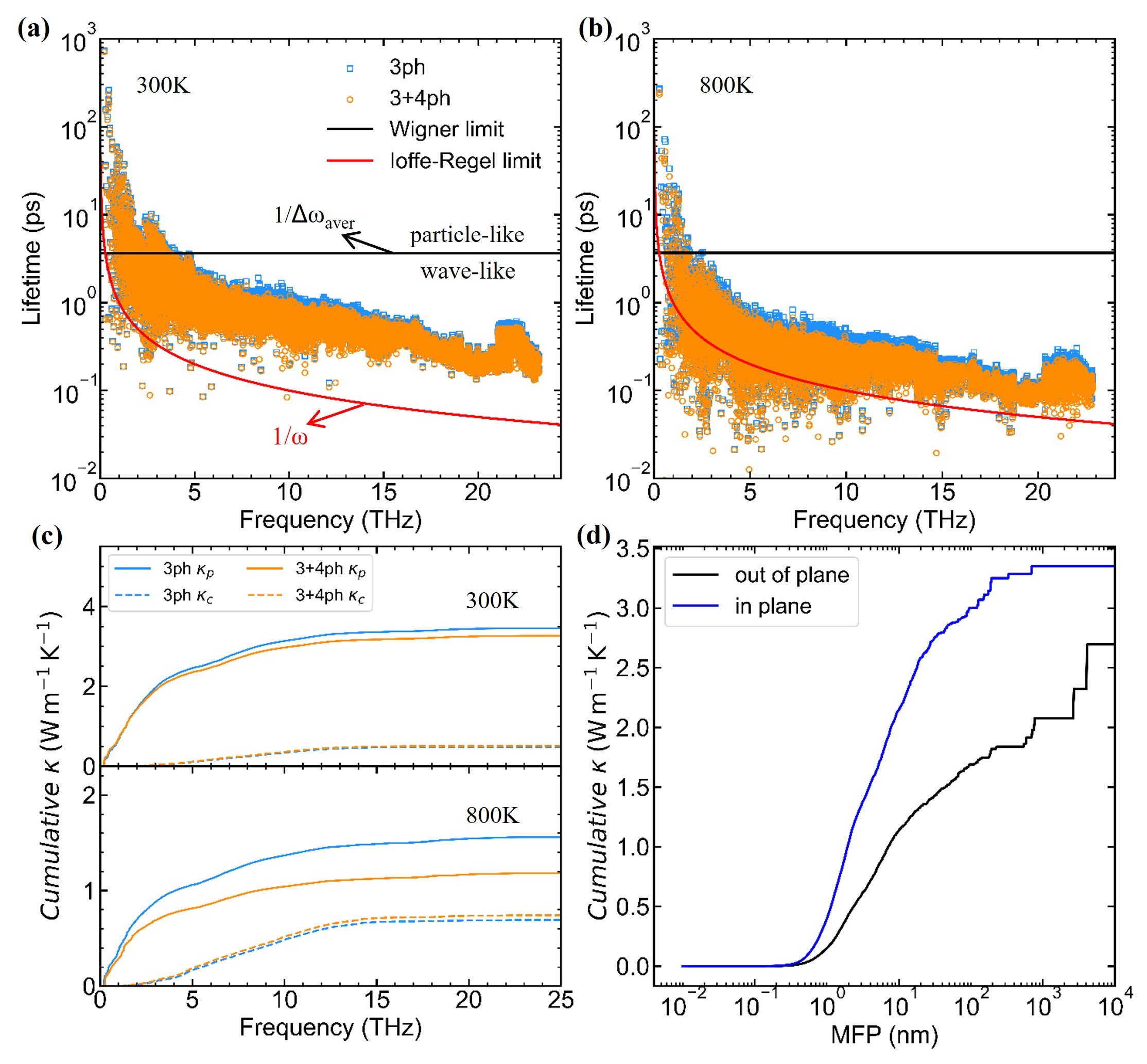} % 适当调整宽度
\caption{
(a) Comparison of phonon lifetimes due to three-phonon and four-phonon scattering processes at 300 K.  (b) Same as panel (a), but at 800 K. (c) Cumulative thermal conductivity due to three-phonon and four-phonon processes, decomposed into contributions from diffusons and propagating phonons. Top panel: 300 K; bottom panel: 800 K. (d) Cumulative MFP thermal conductivity at room temperature.
}
\label{fig4}
\end{figure*}

\subsection{Two-channel thermal conductivity}
Given its complex unit cell, $\rm LiYTiO_4$ exhibits intricate phonon band structure with the large number of closely spaced optical phonon branches (Fig.~\ref{fig2}(c)), along with strong anharmonicity as revealed by its large Grüneisen parameters (see the Supplementary Material). These features collectively signal that phonon wave-particle duality plays a critical role in its thermal transport. To verify this inference, we plot in Fig.~\ref{fig4}(a) the computed phonon lifetimes at 300 K against the Ioffe-Regel limit in time \cite{C3TA15147F} (corresponding to a phonon scattering rate equal to the phonon frequency, i.e., $1/\tau_{\rm Ioffe-Regel} =\omega$) and the Wigner limit \cite{simoncelli_unified_2019} (corresponding to a phonon scattering rate equal to the average interband spacing, i.e., $1/\tau_{\rm Wigner}=\Delta\omega_{\rm ave}$). Phonons with $\tau$ longer than the Wigner limit contribute to $\kappa_{\mathrm{L}}^{\rm P}$ via particle-like propagation, while those with scattering rates between the Ioffe-Regel and Wigner limits contribute to $\kappa_{\mathrm{L}}^{\rm C}$ via wave-like tunneling. Critically, our results show that the majority of phonons reside in this wave-tunneling regime, a trend that becomes even more pronounced when 4ph scattering is included, underscoring the pivotal role of coherent transport. It is clearly seen from Fig.~\ref{fig4}(d) that when the temperature increases to 800 K, a greater number of phonons drop below the Wigner limit, indicating that the contribution of wave-like tunneling to $\kappa_{\rm L}$ grows significantly with temperature.

% 0.378533 ÷ 1.558766 ≈ 20.51%
With the finite-temperature IFCs obtained, we proceed to calculate the $\kappa_{\rm L}$ of $\rm LiYTiO_4$ using the Wigner transport equation, accounting for both phonon population and coherence contributions. It is evident from the cumulative $\kappa_{\rm L}$ with frequency in Fig.~\ref{fig4}(c) that while particle-like transport dominates $\kappa_{\rm L}$ from 300 to 800 K, wave-like coherence also plays a substantial role, especially at high temperatures. Notably, phonons below 15 THz account for the majority of both transport contributions. Furthermore, we find that 4ph scattering significantly suppresses the particle-like thermal conductivity at high temperatures, reducing $\kappa_{\mathrm{L}}^{\rm P}$ by approximately 20.51\% at 800 K, while having a negligible effect on the wave-like tunneling contribution.

From an experimental perspective, the phonon mean free path (MFP) spectra can provide crucial insights into the influence of grain size on thermal conductivity. As shown in Fig.~\ref{fig4}(d), the dominant phonon MFP along the in-plane and out-of-plane directions in $\rm LiYTiO_4$ at 300 K is below 1$\mu$m, and this value is expected to decrease further at elevated temperatures. This finding confirms that in our synthesized polycrystalline samples, lattice thermal conductivity cannot be affected by grain boundaries. Therefore, a direct comparison between our theoretical prediction and experimental measurements is well justified.

\begin{figure*}[htbp]
\centering
\includegraphics[width=0.8\linewidth]{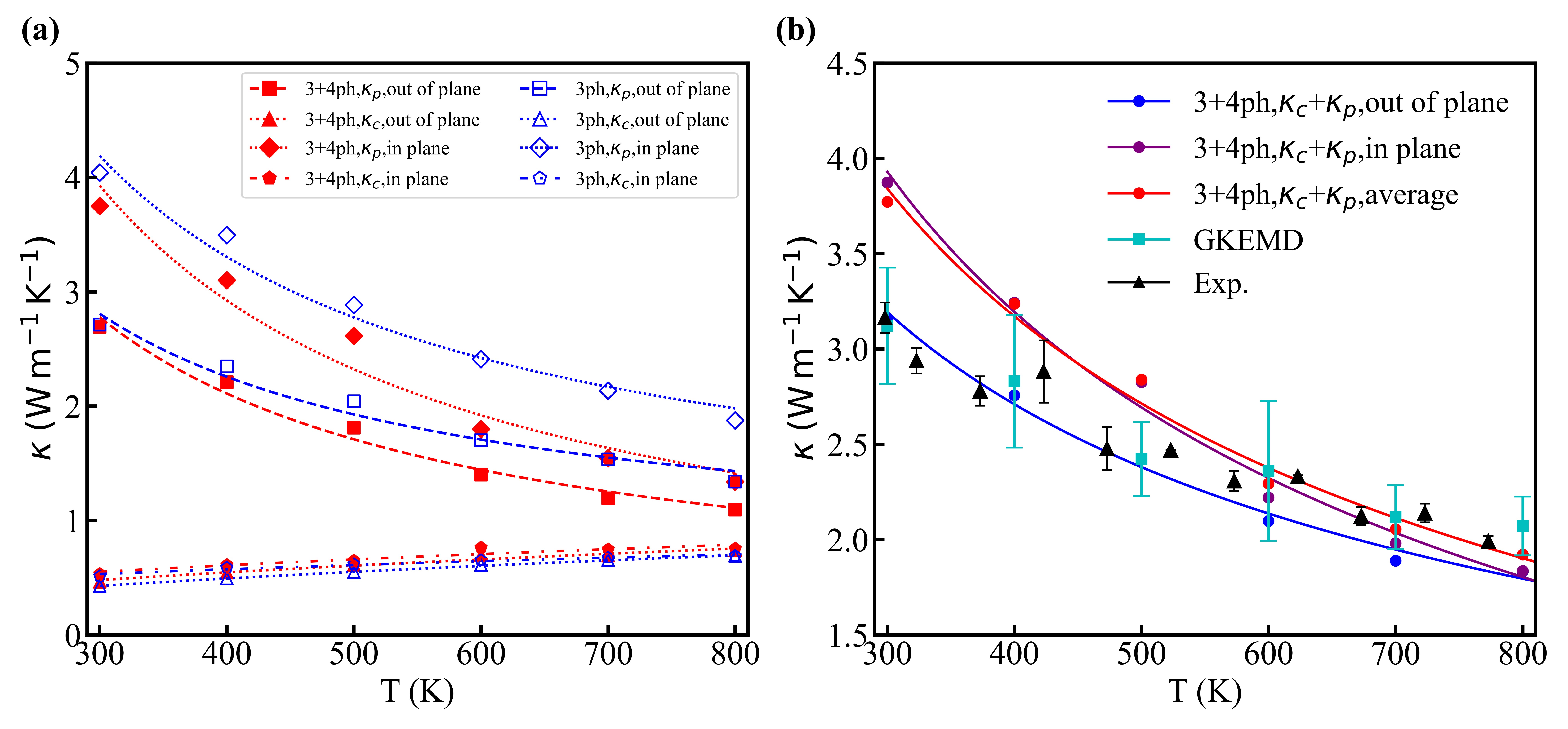} % 适当调整宽度
\caption{
(a) In-plane and out-of-plane thermal conductivity due to three-phonon and four-phonon scattering processes, decomposed into contributions from diffusons and propagating phonons.  (b) Comparison of in-plane, out-of-plane, and averaged thermal conductivities with GKEMD simulations and experimental measurements.
}
\label{fig5}
\end{figure*}

Figure~\ref{fig5}(a) shows the calculated temperature-dependent thermal conductivities from the Wigner formalism for the coherent ($\kappa_{\mathrm{L}}^{\rm C}$ ) and particle ($\kappa_{\mathrm{L}}^{\rm P}$) contributions along the in-plane and out-of-plane directions. It can be observed that the in-plane particle-like conductivity ($\kappa_{\mathrm{L}}^{\rm P}$) is significantly larger than its out-of-plane counterpart at lower temperatures, regardless of whether 4ph scattering is included, whereas the difference in their coherent components ($\kappa_{\mathrm{L}}^{\rm C}$) is minimal. This pronounced anisotropy in thermal conductivity originates primarily from the layered crystal structure of $\rm LiYTiO_4$, which features considerably weaker chemical bonding within the planes.  

To validate our theoretically predicted lattice thermal conductivity of $\rm LiYTiO_4$, we further compare the calculated total values ($\kappa_{\mathrm{L}}=\kappa_{\mathrm{L}}^{\rm P}+\kappa_{\mathrm{L}}^{\rm C}$) with our experimental measurements over the temperature range of 300–800 K, as presented in Fig.~\ref{fig5}(b). Given that the experiments were performed on polycrystalline samples, the theoretical values are averaged over the in-plane and out-of-plane directions for a direct comparison. It is seen that although the Wigner formalism does not explicitly include the effect of ion hopping, the predicted thermal conductivity still exhibits reasonable agreement with the experimental data across the entire temperature range. This suggests that ion hopping has only a minor influence on thermal transport in $\rm LiYTiO_4$. The negligible influence of ion hopping can be understood by its resemblance to the rattling-like motion found in certain superionic crystalline materials \cite{article, PhysRevLett.120.105901, PRXEnergy.4.013012}, which introduces strong anharmonicity. This anharmonicity strongly suppresses particle-like transport while concurrently enhancing the wave-like coherence contribution. As a result, within the two-channel thermal transport regime, ion hopping contributes very little to the total thermal conductivity. A similar phenomenon has also been reported in other liquid-like systems such as $\rm Cu_3BiS_3$ \cite{https://doi.org/10.1002/smll.202506386}. In comparison, the total thermal conductivity predicted by the GK-EMD method closely matches the experimental data over the entire temperature range. This agreement confirms the superior reliability of machine-learning-based MD simulations in predicting thermal transport in superionic crystals like $\rm LiYTiO_4$, as they inherently incorporate all-order phonon scattering and the contributions from ionic hopping.

Combining theoretical predictions with experimental results, we find that the upper limits of the thermal conductivity of $\rm LiYTiO_4$ along the in-plane and out-of-plane directions are only 4.28 and 3.16 W/mK at 300K, respectively, which are significantly lower than those of conventional layered lithium transition-metal oxides (LiTMO$_2$) \cite{FENG2020104916}. This implies that despite its outstanding electrochemical performance, the ultralow thermal conductivity of $\rm LiYTiO_4$ may constitute a major bottleneck to its practical application.

\section{CONCLUSION}
In summary, we have established a fundamental understanding of the lattice dynamics and thermal transport in $\rm LiYTiO_4$ through a combined experimental and theoretical study. Phonon calculations at 0 K reveal dynamical instabilities arising from opposite rotations of adjacent $\rm TiO_6$ octahedra, which are stabilized at finite temperatures via anharmonic renormalization. By employing the WTT formalism, which explicitly incorporates 3ph and 4ph scattering processes and accounts for both particle-like and wave-like contributions, we predict that orientation-averaged $\rm κ_L$ is 3.8 $\rm Wm^{-1}K^{-1}$ at room temperature. This value is in good agreement with the experimentally measured $\kappa_{\rm L}$ of 3.2 ± 0.08 $\rm Wm^{-1}K^{-1}$ for polycrystalline samples. The MSD calculations further show that Li‑ion diffusion initiates above 600 K and becomes more pronounced with rising temperature. To evaluate the possible role of such ionic hopping on high‑temperature thermal transport, we also compute $\kappa_{\rm L}$ using a NEP-based GK-EMD method. The results are reasonably consistent with both experimental measurements and WTT predictions, demonstrating that Li-ion mobility plays a negligible role in determining $\kappa_{\rm L}$ eve at high temperatures. Our findings clarify the limited contribution of ionic diffusion to heat transport in $\rm LiYTiO_4$. Moreover, the ultralow $\kappa_{\rm L}$ identified here underscores thermal management as a critical challenge for the practical application of $\rm LiYTiO_4$ in lithium‑ion battery anodes.

\textit{Acknowledgements.} This work is supported by the National Natural Science Foundation of China (Grant No. 12374038 ) and the Fundamental Research Funds for the Central Universities (Grant No. 2025CDJ-IAISYB-035). 

%\nocite{*}   % 把 .bib 文件中的所有文献都放进参考文献列表，即使在正文中没有被 \cite{} 引用


\end{document}

\begin{center}
Supplementary information for
\end{center}

%\title{Significant reduction of the lattice thermal conductivity in boron arsenide by symmetry-breaking strain}
\title{Probing Anharmonic Lattice Dynamics and Thermal Transport in Fast-Charging Lithium-Ion Batteries: A Case Study of Layered Perovskite}
\author{Lin Zhang}
\affiliation{College of Physics and Center of Quantum Materials and Devices, Chongqing University, Chongqing 401331, China}
\affiliation{College of Physics, Chongqing Key Laboratory for Strongly Coupled Physics, Chongqing University, Chongqing 401331, China}

\author{Jun Huang}%
\email{huangjun2003@126.com}
\affiliation{School of Chemistry and Chemical Engineering, Frontiers Science Center for Transformative Molecules, Shanghai Jiao Tong University, Shanghai 200240, China}

\author{Xiaolong Yang}
\email{yangxl@cqu.edu.cn}
\affiliation{College of Physics and Center of Quantum Materials and Devices, Chongqing University, Chongqing 401331, China}
\affiliation{College of Physics, Chongqing Key Laboratory for Strongly Coupled Physics, Chongqing University, Chongqing 401331, China}

%\thanks{*Corresponding author: yangxl@cqu.edu.cn}

\date{\today}

%\keywords{}

\maketitle

\section{S1. Machine Learning Potential}
\vspace{-3cm}
\begin{figure}[!ht]
\includegraphics[width=\columnwidth]{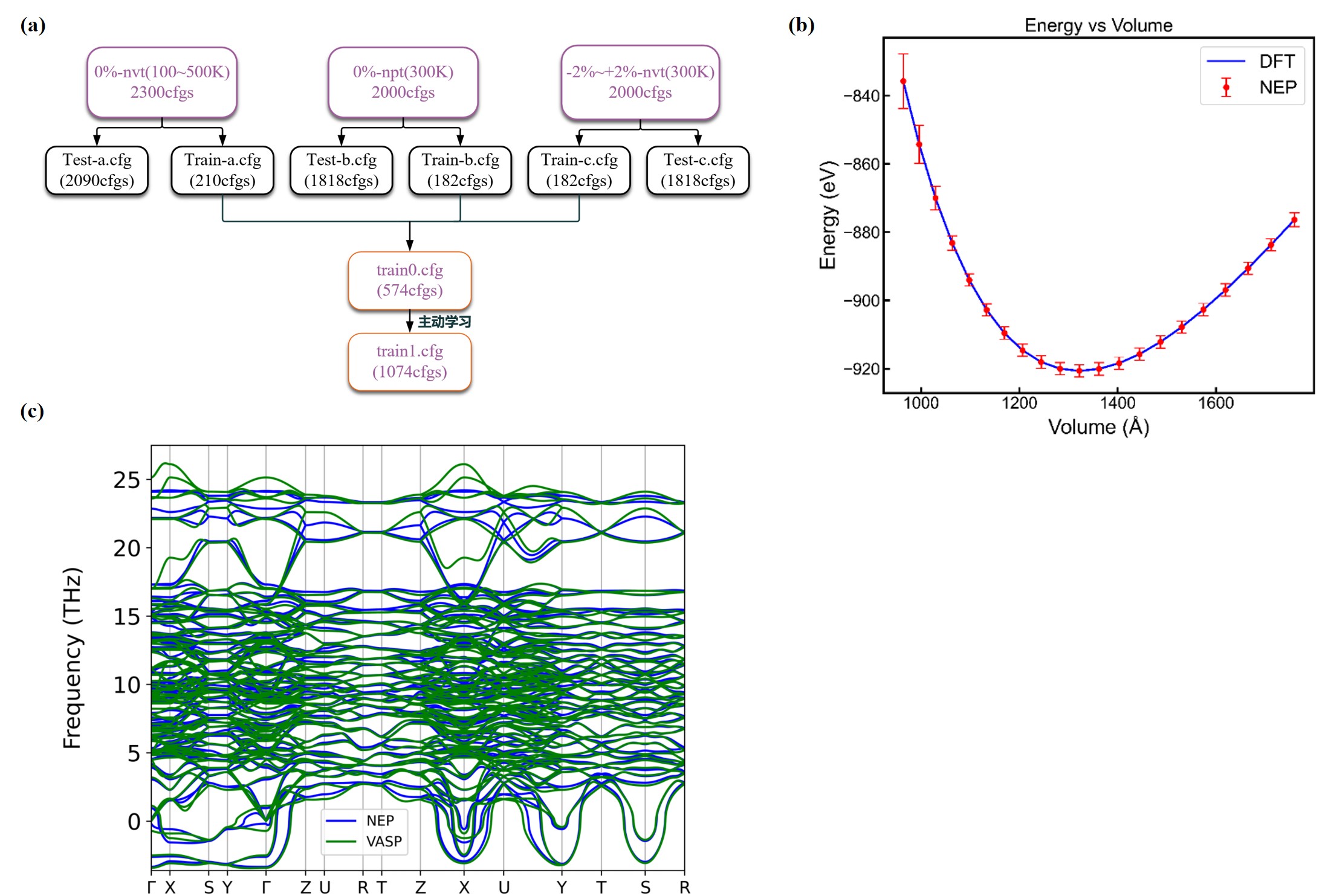}%
\caption{\label{phdisp} (a) Composition of the training set and training workflow.
(b) Relationship between the energy of the supercell and volume, obtained from NEP and DFT, respectively.
(c) Comparison of the phonon spectra obtained from NEP and DFT.}
\end{figure}

\section{S2. Manifestation of Anharmonicity }
\begin{figure}[H]
\centering
\includegraphics[scale=0.8]{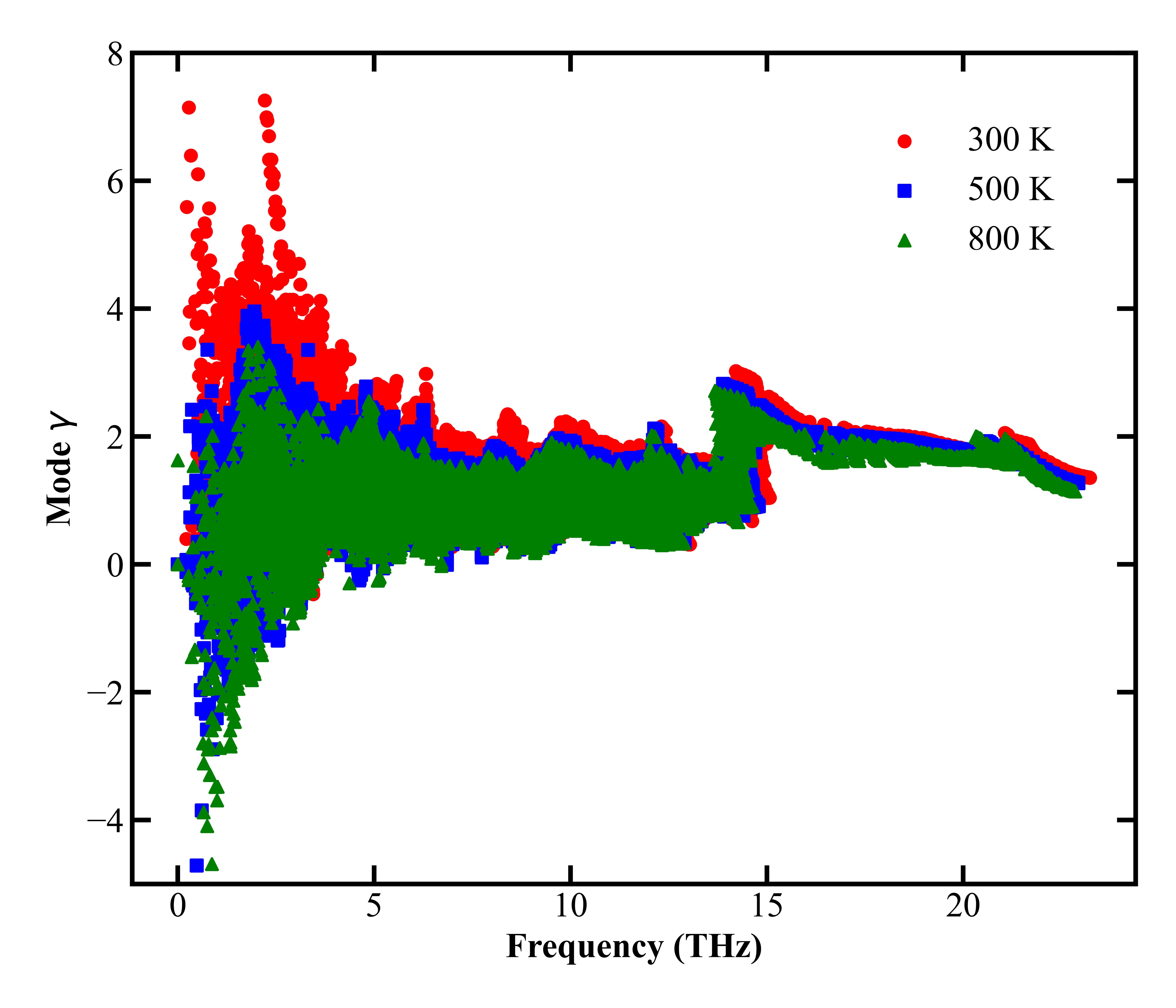}
\caption{\label{phdisp} Mode Grüneisen parameters of $LiYTiO_4$ at 300K, 500K, 800K. }
\end{figure}

\section{S3. Convergence Tests for Shengbte}
\begin{figure}[H]
\includegraphics[width=\columnwidth]{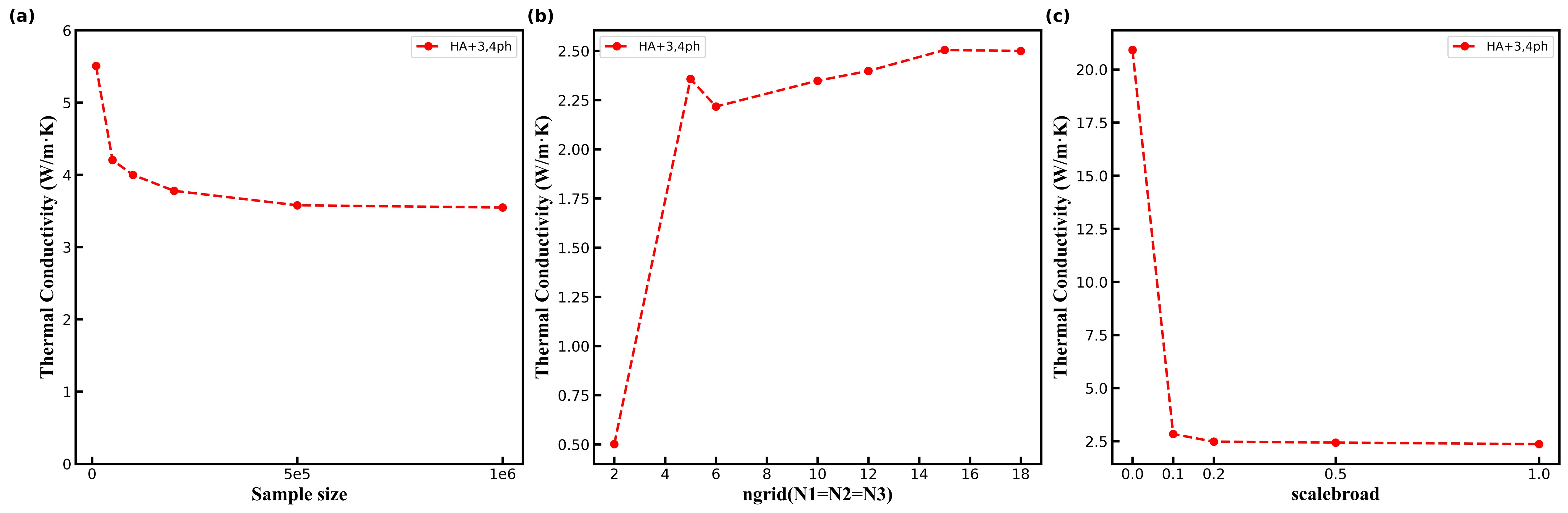}%
\caption{\label{phdisp} Thermal conductivity convergence tests:
(a) Four-phonon sampling density tests at 300 K.
(b) q-point mesh density tests at 500 K.
(c) Gaussian broadening value (scalebroad) tests at 500 K.}
\end{figure}

\section{S4. Characterization of Ion Diffusion }
%\vspace{-1cm}
\begin{figure}[H]
%\vspace{-150pt}   % 手动压缩距离
\includegraphics[width=\columnwidth]{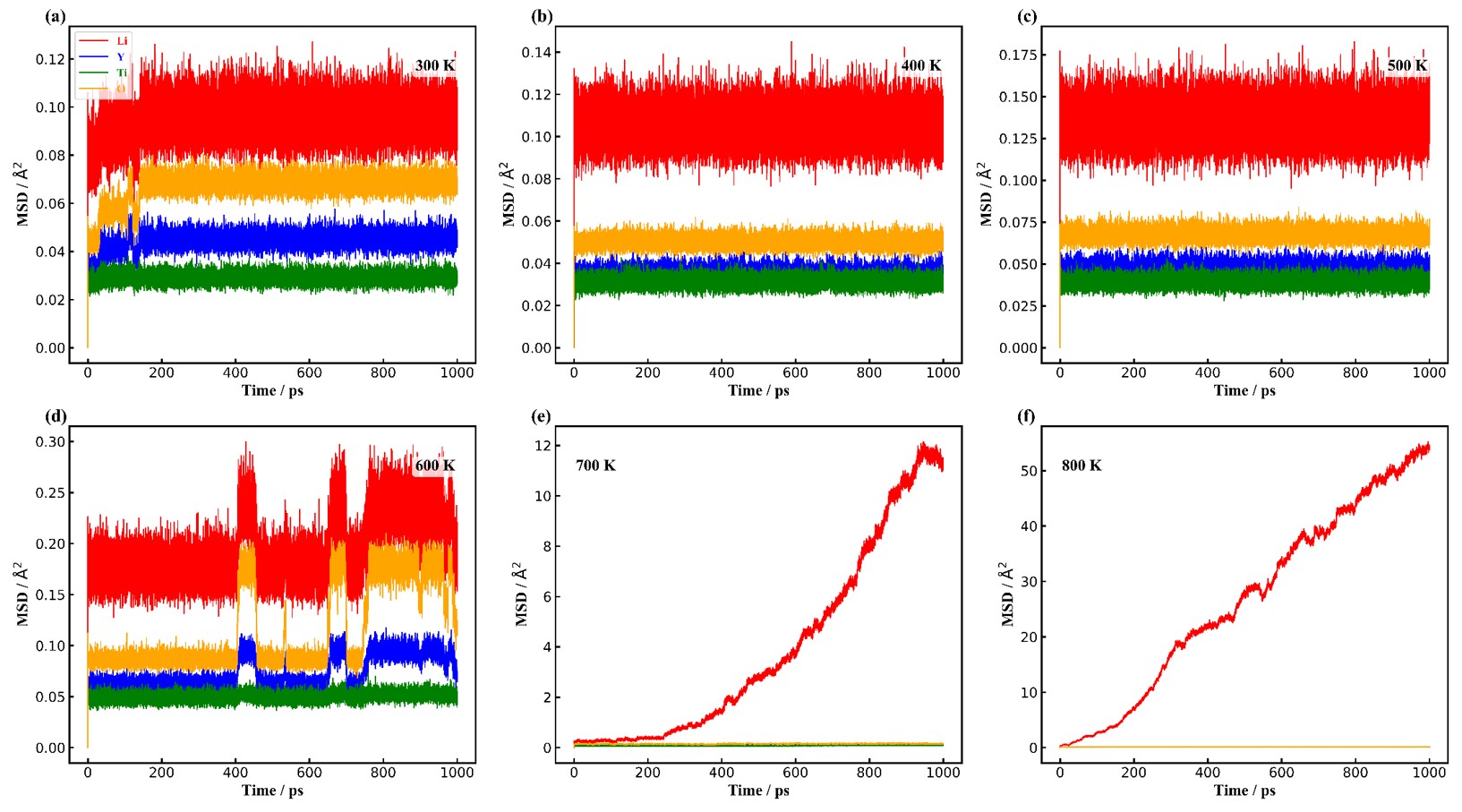}%
\caption{\label{phdisp} Mean square displacements (MSDs) of Li, Y, Ti, and O atoms in LYTO from 300 K to 800 K.}
\end{figure}

\section{S5. Synthesis of LiYTiO\_4 Samples }
\begin{figure}[H]
\includegraphics[width=\columnwidth]{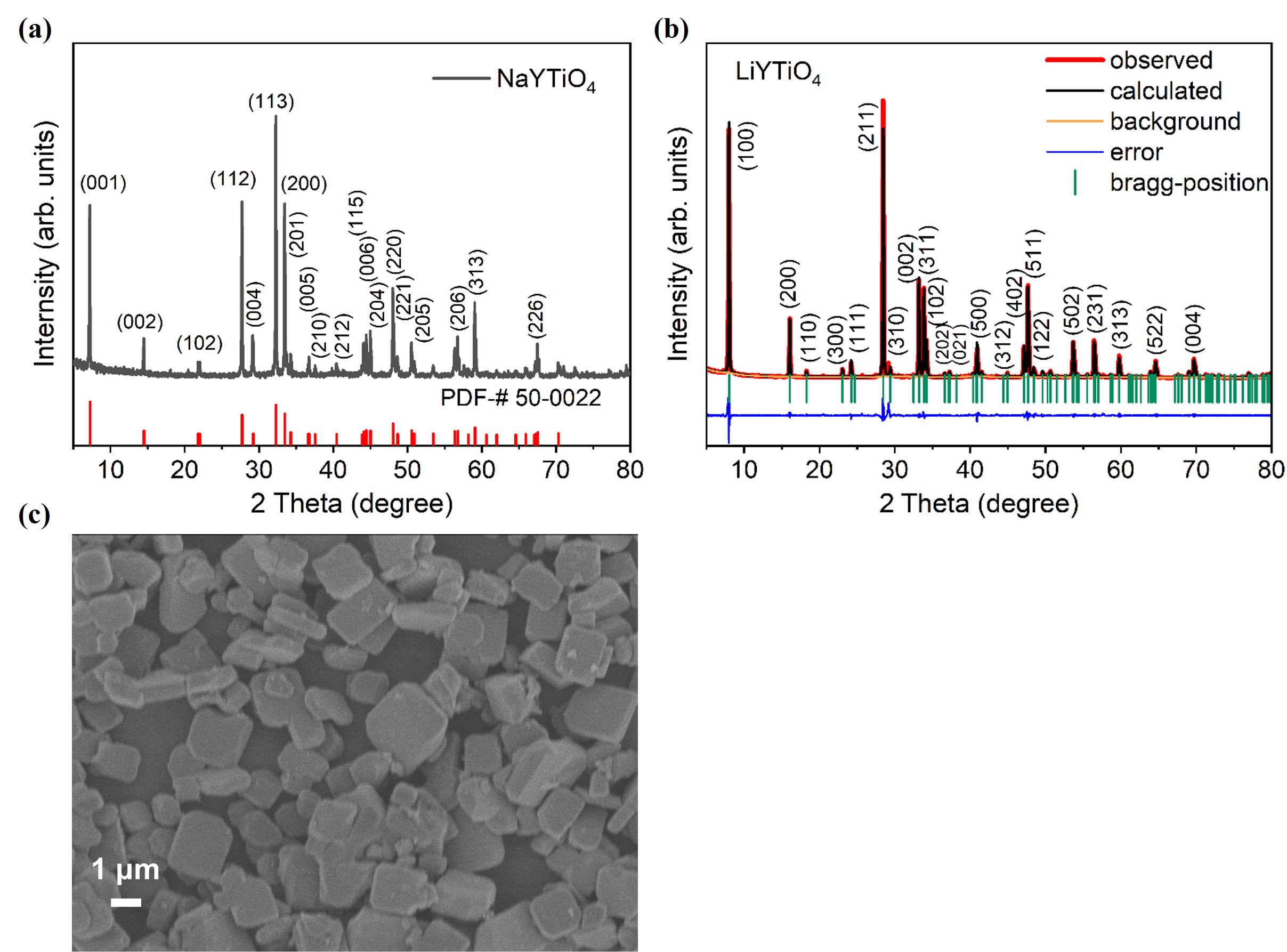}  %
\caption{\label{phdisp} (a)	XRD pattern of NaYTiO4; (b) Rietveld refinement XRD pattern of LiYTiO4 (c) SEM image of LiYTiO4.}
\end{figure}

%This study employed a sol--gel method combined with an ion-exchange process to synthesize the compound $\mathrm{LiYTiO_4}$. The synthesis first involved preparing a sodium-based precursor, $\mathrm{NaYTiO_4}$, via the sol--gel method, followed by a lithium--sodium ion-exchange reaction. Compared with the conventional high-temperature solid-state calcination method (which typically requires prolonged high-temperature treatment), this synthesis strategy significantly shortens the calcination duration and effectively suppresses phase separation and abnormal grain growth that may occur at elevated temperatures.

%$\mathrm{LiYTiO_4}$ crystallizes in an orthorhombic structure with space group \textit{Pbcm} (No.~57). The refined %lattice parameters are 
%\[
%a = 11.0196(5)\,\text{\AA}, \quad 
%b = 5.3994(4)\,\text{\AA}, \quad 
%c = 5.3929(5)\,\text{\AA}.
%\]
%As illustrated in Fig.~2(a), the crystal structure of $\mathrm{LiYTiO_4}$ exhibits a layered arrangement along the $a$-axis, consisting of alternating layers of $\mathrm{LiO_4}$ tetrahedra, $\mathrm{YO_8}$ dodecahedra, and $\mathrm{TiO_6}$ octahedra. Due to the difficulty of preparing high-quality single crystals, polycrystalline $\mathrm{LiYTiO_4}$ samples with grain sizes of approximately $1$--$2\,\mu\mathrm{m}$ were synthesized in this study.

\bibliographystyle{iopart-num}
%\bibliography{common}